\documentclass[12pt]{iopart}
\input{amssym}
\usepackage{epsfig}

\global\arraycolsep=2pt

\begin{document}
\title[Drude Model
and Casimir Entropy]{Analytical and Numerical Demonstration of How the
Drude Dispersive Model Satisfies Nernst's Theorem for the Casimir Entropy}

\author{Iver Brevik$^1$, Simen A. Ellingsen$^1$,
Johan S. H{\o}ye$^2$ and Kimball A. Milton$^3$}

\address{$^1$ Department of Energy and Process Engineering,
Norwegian University of Science and Technology, N-7491 Trondheim, Norway\\
$^2$ Department of Physics, Norwegian University of Science and
Technology, N-7491 Trondheim, Norway\\
$^3$ Homer L. Dodge Department of Physics and Astronomy, The
University of Oklahoma, Norman, OK 73019, USA}

\ead{iver.h.brevik@ntnu.no}

Revised version, \today

\begin{abstract}
In view of the current discussion on the subject, an effort is
made to show very accurately both analytically and numerically how
the Drude dispersion model, assuming the relaxation is nonzero at
zero temperature (which is the case when impurities are present),
gives consistent results for the Casimir free energy at low
temperatures. Specifically, we find that the free energy consists
essentially of two terms, one leading term proportional to  $T^2$,
and a next term proportional to $T^{5/2}$. Both these terms give
rise to zero Casimir entropy as $T \rightarrow 0$, thus in
accordance with Nernst's theorem.
\end{abstract}


\section{Introduction}

\begin{figure}[h]
\begin{center}
\includegraphics[width=3.5in]{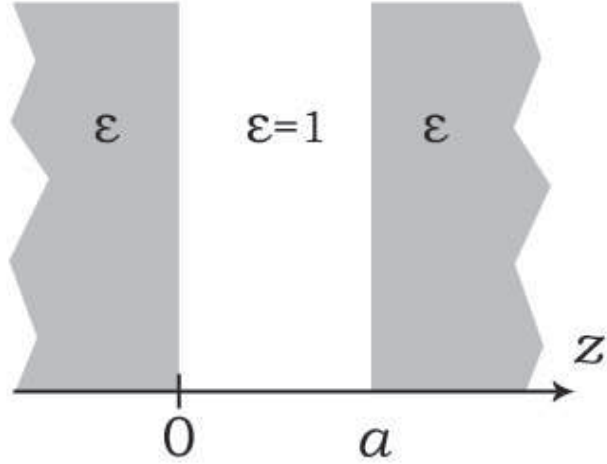}
\caption{\label{fig1}Sketch of the geometry}
\end{center}
\end{figure}

The thermodynamic consistency of the expression for the Casimir
pressure at finite temperature $T$ is of considerable current
interest. The problem gets accentuated at low $T$, where according
to Nernst's theorem $S=-\partial F/\partial T \rightarrow 0$ when
$T\rightarrow 0$. Here $S$ is the entropy and $F$ the free energy per
unit surface area. We shall consider the standard Casimir
configuration, namely two semi-infinite identical metallic media
separated by a vacuum gap of width $a$. The media are assumed
nonmagnetic with a frequency-dependent  relative permittivity
$\varepsilon(\omega)$. The two surfaces lying at $z=0$ and $z=a$
are taken to be perfectly planar and of infinite extent. A sketch
of the setup is given in \fref{fig1}.

The present work is closely related to our  recent paper \cite{hoye07} in
particular, and also to our earlier papers on the thermal Casimir
effect \cite{brevik06,brevik06a,brevik05,brevik04,hoye03}.

We start from the Lifshitz formula:
\begin{equation}
  \beta
F=\frac{1}{2\pi}{\sum_{m=0}^\infty}{}'\int_{\zeta_m/c}^\infty
\left[\ln(1-A\rme^{-2qa})+\ln(1-B\rme^{-2qa})\right]q\,\rmd q, \label{1}
\end{equation}
where
\numparts
\begin{eqnarray}
  A=\left(\frac{s-\varepsilon p}{s+\varepsilon p}\right)^2
\quad (\mbox{TM mode}) \label{2}
\\
 B=\left(\frac{s-p}{s+p}\right)^2 \quad (\mbox{TE mode}) \label{3}
\\
 \zeta_m=\frac{2\pi k}{\hbar}mT, \quad \beta=1/kT \label{4}
 \\
  s=\sqrt{\varepsilon-1+p^2}, \quad p=\frac{qc}{\zeta_m}. \label{5}
\end{eqnarray}
\endnumparts
Here $ \zeta_m$ are the Matsubara frequencies, $s$ and $p$ are the
Lifshitz variables, and the prime on the summation sign means that
the case $m=0$ is to be taken with half weight.

The appropriate dispersion relation to use is the Drude relation
\begin{equation}
 \varepsilon(i\zeta)=1+\frac{\omega_p^2}{\zeta(\zeta+\nu)},\label{6}
 \end{equation}
 where $\omega =i\zeta$, $\omega_p$ being the plasma frequency, and
 $\nu$ the relaxation frequency. The plasma wavelength is
 $\lambda_p=2\pi c/\omega_p$. Our motivation for adopting the form
 (\ref{6}) is that it agrees well with permittivity measurements
 (performed at room temperature). In the case of gold,
\begin{equation}\,
 \omega_p=9.03 \,{\rm eV},~
\nu=34.5\, {\rm meV},~ \lambda_p=137.4\,
{\rm nm}.\label{7}
\end{equation}
The  Drude relation is good for $\zeta< 2\times 10^{15}$ rad/s.
For higher $\zeta$, the relation gives too low values for the
permittivity (cf. figure 1 in \cite{hoye03}). Actually, the
numerical input data we used were taken directly
from tabulated data along the imaginary frequency axis,
$\varepsilon(i\zeta)$ for $\zeta >0$ (courtesy of Astrid
Lambrecht). These data cover 7 decades:
\begin{equation}
 1.5\times 10^{11}<\zeta<1.5\times 10^{18} ~{\rm rad/s}.\label{8}
 \end{equation}
For $\zeta<1.5\times10^{11}$ values for $\varepsilon(i\zeta)$ are
obtained from (\ref{6})
 by extrapolation, but by our numerical evaluations only the $m=0$ value fell within this region.

 As mentioned, the permittivity measurements are made at room
 temperature. For definiteness, we shall in the following use the
 room-temperature value $\nu= 34.5 \,$ meV already given in (\ref{7}),
 although  we expect that at very low temperatures the true
 value of $\nu$ is actually lower - cf. the recent discussion on
 this point by Klimchitskaya and Mostepanenko
 \cite{klimchitskaya07}. This fact will change our results
 quantitatively, but not qualitatively. In particular, it will not
 change our main conclusion regarding the validity of the Nernst
 theorem when $T\rightarrow 0$.

 Let us emphasize the main assumption underlying our calculations:
 {\it We assume $\nu$ to possesses a nonzero
 value, however small, at any fixed temperature including $T=0$.}

 The assumed constancy of $\nu$ might be questioned, as the Bloch-Gr{\"u}neisen law predicts that $\nu$
 depends on $T$ as (cf. Appendix D in \cite{hoye03})

 \begin{equation}
 \nu(T)\propto T^5, \quad T\rightarrow 0.
\end{equation}
Such a relationship is not followed in practice, however, since
there are always impurities which give rise to nonzero resistivity
and so nonzero relaxation frequency at zero temperature
\cite{khoshenevisan79}. In practice, therefore, our assumption
above is always satisfied. The important point is that the
relationship
\begin{equation}
 \zeta^2[\varepsilon(i\zeta)-1]\rightarrow 0,\quad\zeta
\rightarrow 0 \label{11}
\end{equation}
is always satisfied. It implies that the zero-frequency
TE mode does not contribute to the Casimir force. The first to
emphasize this kind of behaviour were Bostr{\"o}m and Sernelius
\cite{bostrom00}, and the issue was discussed in detail in
\cite{hoye03}. There are several other papers arguing along
similar lines. Thus Jancovici and \v{S}amaj \cite{jancovici00} and
Buenzli and Martin \cite{buenzli05} considered  the classical
plasma of free charges in the
high-temperature limit, where only zero frequency contributes,
and they found the linear dependence in T in the Casimir
force to be reduced by a factor of 2 from the behaviour of an {\it
ideal} metal (the IM model).


To illustrate the magnitude of the Drude thermal correction to the
Casimir pressure, we give in Table 1 some calculated values, in
mPa. It should be noted that if $T$ increases from 300 K to 350 K,
we find that

(i) if  $a=0.2 ~\mu$m, the Casimir pressure diminishes by 0.4\%;

(ii) if $a=2.0~\mu$m, the  Casimir pressure diminishes by 3.7\%.

 The optimum gap width in connection with Casimir thermal
 corrections thus seem to lie around $a=2\,\mu$m.

\begin{table}[h]
\caption{The Casimir pressure (in mPa) between Au-Au plates versus
gap width $a$, when $T=\{ 1,300,350\}$ K. Data extracted from
Ref.~\cite{brevik06a}.}
\begin{indented}
\item[]\begin{tabular}{cccc}
$a/\mu$m         &   $T=1$ K   &  $T=300$ K  & $T=350$ K\\ \hline \hline

 0.2          &   508.2       &  497.8     &   495.7     \\

 0.5           & 16.56       & 15.49       & 15.30     \\

 1.0           & 1.143       & 0.9852      & 0.9590  \\

 2.0           & $7.549\times 10^{-2}$   & $5.550\times 10^{-2}$     & $5.344\times 10^{-2}$ \\

 3.0           &$1.520\times10^{-2}$     & $ 1.033\times 10^{-2}$    &$1.049\times 10^{-2}$    \\

 4.0           &$ 4.858\times 10^{-3}$     &$3.481\times 10^{-3}$     &$3.804\times 10^{-3}$     \\
\hline \hline
\end{tabular}
\end{indented}
\end{table}

An argument that has been put forward against the Drude relation
is that by omitting the zero frequency TE term  one gets a term
linear in $T$ in the free energy. Such a term would lead to a
finite entropy at $T=0$ and so come into conflict with Nernst's
theorem. There are several recent papers arguing along these
lines, written from somewhat different perspectives
\cite{klimchitskaya07,geyer07,chen07,klimchitskaya07a,bezerra06,decca05,bezerra04},
and it is argued there that for such lattices the Drude model
violates the Nernst theorem. Perfect crystal lattices, without
impurities at all, are considered in these papers. It is argued
therein that for thermodynamical consistency, relaxation due to
electron-phonon scattering present at finite temperature should be
neglected, and the use of the plasma dispersion relation
\begin{equation}
   \varepsilon(i\zeta)=1+\frac{\omega_p^2}{\zeta^2}, \label{12}
   \end{equation}
or a generalized version thereof is presented. The relation
(\ref{12})
   corresponds to setting $\nu=0$ in (\ref{6}),
 such that (\ref{12}) does {\it not} satisfy the condition
   (\ref{11}).
 The plasma relation leads to quite a small
   temperature dependence in the Casimir force (correction $\propto T^4$)
in contrast to the distinct and almost linear decay with the Drude
relation. Actually, the Drude theory in the limit $\nu \rightarrow
0$  preserves entropy $S=0$ at $T=0$, but $S$ changes more and
more abruptly at $T=0$ the smaller $\nu$ is.

   In the following we intend to show
   very accurately, both analytically and numerically, how the
   Drude relation with $\nu \neq 0$ leads to results that are in full agreement with
   the Nernst theorem.

\section{Analytical approach}

We start from the Drude model  assuming some constant value for
$\nu $, and consider in the following only Matsubara frequencies
that are relatively small, $\zeta\, (\equiv \zeta_m) \ll \nu$.
These frequencies are the crucial ones for the behavior in the
$T\rightarrow 0$ limit. It is always possible to consider these
frequencies when $\nu$, as mentioned above, is finite. Then,
\begin{equation}
 \varepsilon(i\zeta)=\frac{\omega_p^2}{\zeta(\zeta+\nu)} \approx
\frac{D}{\zeta}, \quad D = \frac{\omega_p^2}{\nu}. \label{13}
\end{equation}
 We consider only the TE mode, which is the mode of main interest.  Replace $q$ by $x$:
 \begin{equation}
  x^2=\frac{q^2c^2}{(\varepsilon
-1)\zeta^2}=\frac{q^2c^2}{D\zeta}, \quad \zeta \ll \nu. \label{14}
\end{equation}
Then the TE mode coefficient (\ref{3}) becomes
\begin{equation}
 B=(\sqrt{1+x^2}-x)^4, \label{15}
\end{equation}
and the TE part of the free energy can be written
\begin{equation}
 \beta F^{TE}=C{\sum_{m=0}^\infty}{}'g(m),\label{16}
 \end{equation}
 where
 \begin{equation}
 g(m)=m\int_{\sqrt{\zeta/D}}^\infty
x\ln\left[1-B\exp{\left(-\frac{2a}{c} \sqrt{D\zeta}\,x\right)} \right]
\rmd x. \label{17}
\end{equation}
Now invoke the Euler-Maclaurin formula:
\begin{equation}
{\sum_{m=0}^\infty}{}'g(m)=\int_0^\infty
g(u)\,\rmd u-\frac{1}{12}g'(0)+\frac{1}{720}g'''(0)-... \label{18}
\end{equation}
One then finds that
\begin{equation}
 g'(0)=\int_0^\infty x\ln(1-B)\,\rmd x=-\frac{1}{4}(2\ln
2-1).\label{19}
\end{equation}
And thereby one gets
\begin{equation}
 \Delta F^{\rm TE}=\frac{C}{48\beta}(2\ln
2-1)=\frac{1}{48}\frac{\omega_p^2}{c^2\hbar \nu}(kT)^2(2\ln
2-1),\label{20}
\end{equation}
valid for $T \ll 0.01$\,K. This result was first given by Milton at
the  QFEXT03 workshop \cite{brevik04}.

Including the leading correction (Euler-Maclaurin summation
starting at $m=1$ instead of at zero), one gets \cite{hoye07}
\begin{equation}
\Delta
F^{\rm TE}=\frac{C}{\beta}\left[-\frac{1}{12}g'(0)\right]
\left[1+0.204\frac{3a\sqrt{2\pi C}}{12g'(0)}+...\right].
\label{21}
 \end{equation}
For gold plates, with $a=1~\mu$m
\begin{equation}
 \Delta F^{\rm TE}=C_1T^2[1-C_2T^{1/2}+...], \label{22}
 \end{equation}
 with
 \begin{equation}
 C_1=5.81\times 10^{-13}~{\rm (J/m^2\,K^2)}, \quad
  C_2=3.03~ {\rm K^{-1/2}}. \label{23}
 \end{equation}
 In order to avoid negative values for $T$ slightly larger than
 0.1 K, it is convenient to introduce the
 Pad{\'e} approximant form
 \begin{equation}
 \Delta F_{th}^{\rm TE}=\frac{C_1T^2}{1+C_2T^{1/2}}. \label{24}
 \end{equation}
 This is equivalent to (\ref{22}) with respect to the first two
 terms. Results (\ref{22}) - (\ref{24}) were first obtained
in Ref.~\cite{hoye07}.

 \section{Numerical calculations}
 In the numerical calculations we assume two gold plates, with $a=1~\mu$m.
 All dispersive data needed are
 in the experimentally known region given by (\ref{8}) above.
As mentioned, the only place where there is a need to use  Drude
relation  (\ref{6}) explicitly, is when $m=0$.  Actually, it is
immaterial whether we use the experimental Lambrecht data
(\ref{8}) or the Drude relation directly. Thus \fref{fig4} is
calculated with the use of the Drude relation for all frequencies,
but it turns out that a practically identical figure (with some
noise because of lower accuracy) is obtained if we use Lambrecht's
data.

At $T=0$ the free energy is calculated numerically as a double
integral rather than a sum of integrals, using a two-dimensional
version of Simpson's method with adaptive quadrature.

As for the TM mode, it is known that for ideal or nonideal metals
the temperature correction for this mode behaves as $T^4$. Thus,
it is a  smaller correction than the $T^2$ and $T^{5/2}$
corrections associated with the TE mode.
We repeat that the dependence of $\nu$ on temperature is
neglected, and that we employ the room-temperature values for
$\nu$ given in (\ref{7}).

\begin{figure}
  \begin{center}
    a)\includegraphics[width=2.5in]{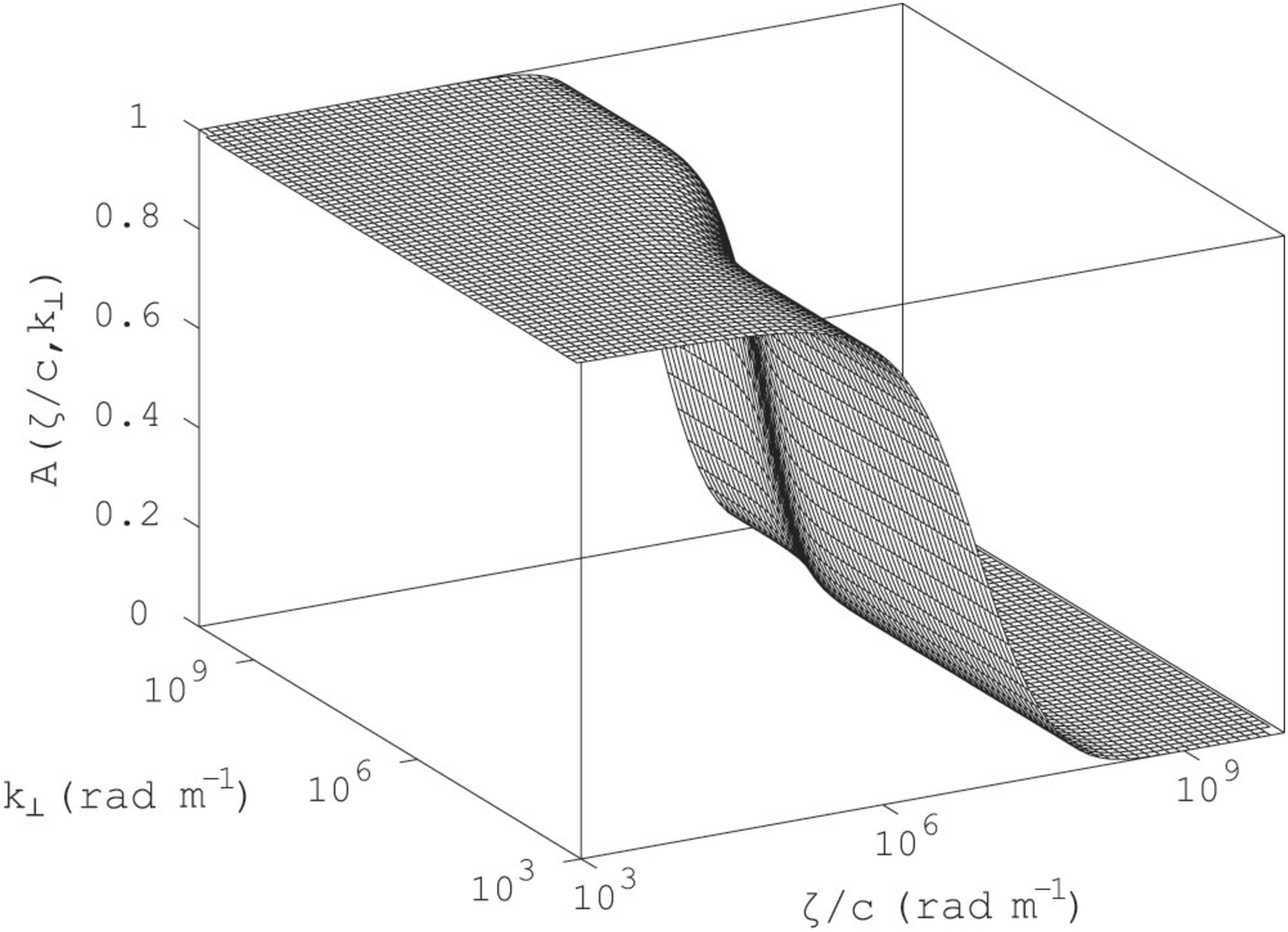}\\
    b)\includegraphics[width=2.5in]{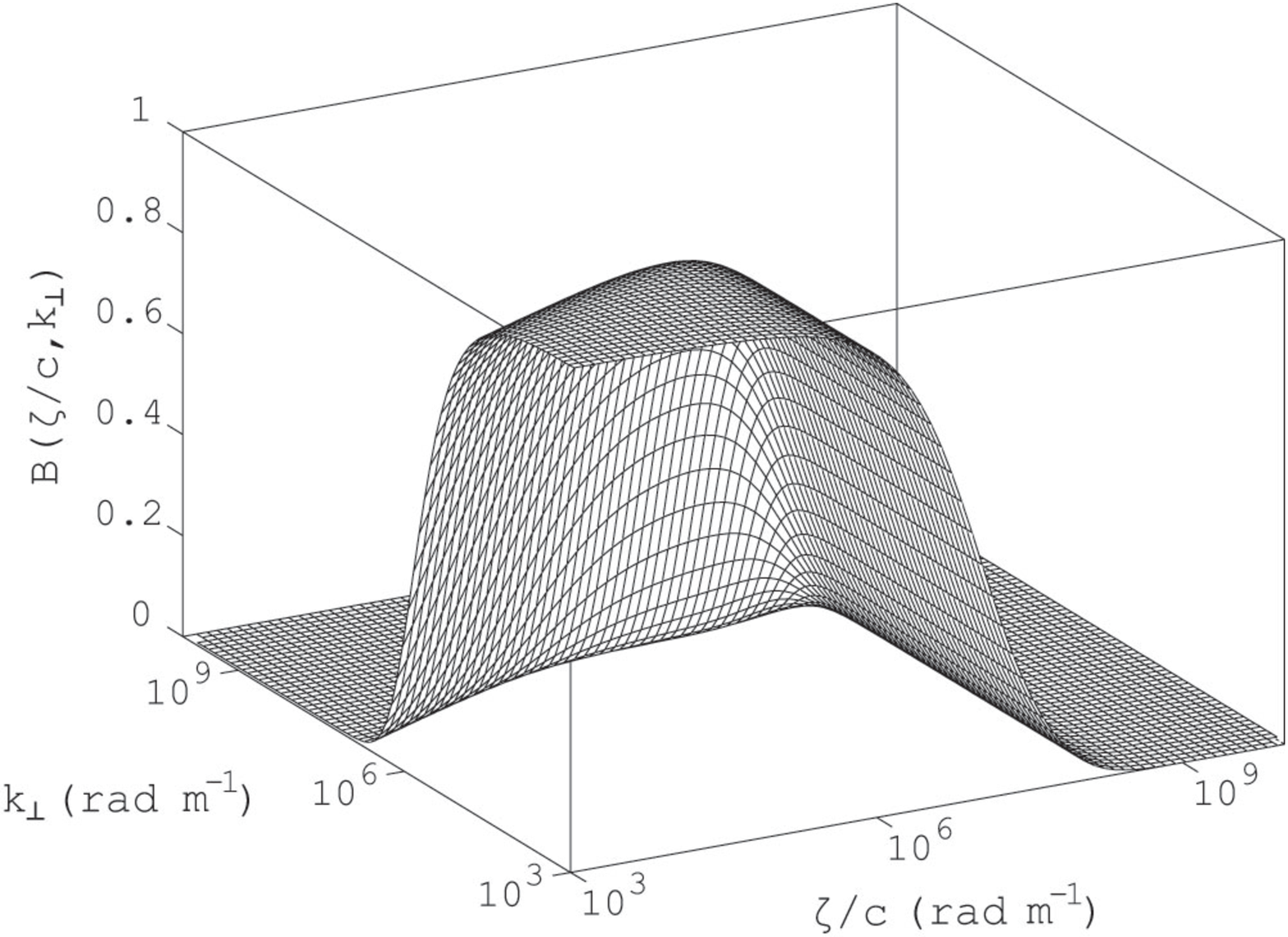}\\
    c)\includegraphics[width=2.5in]{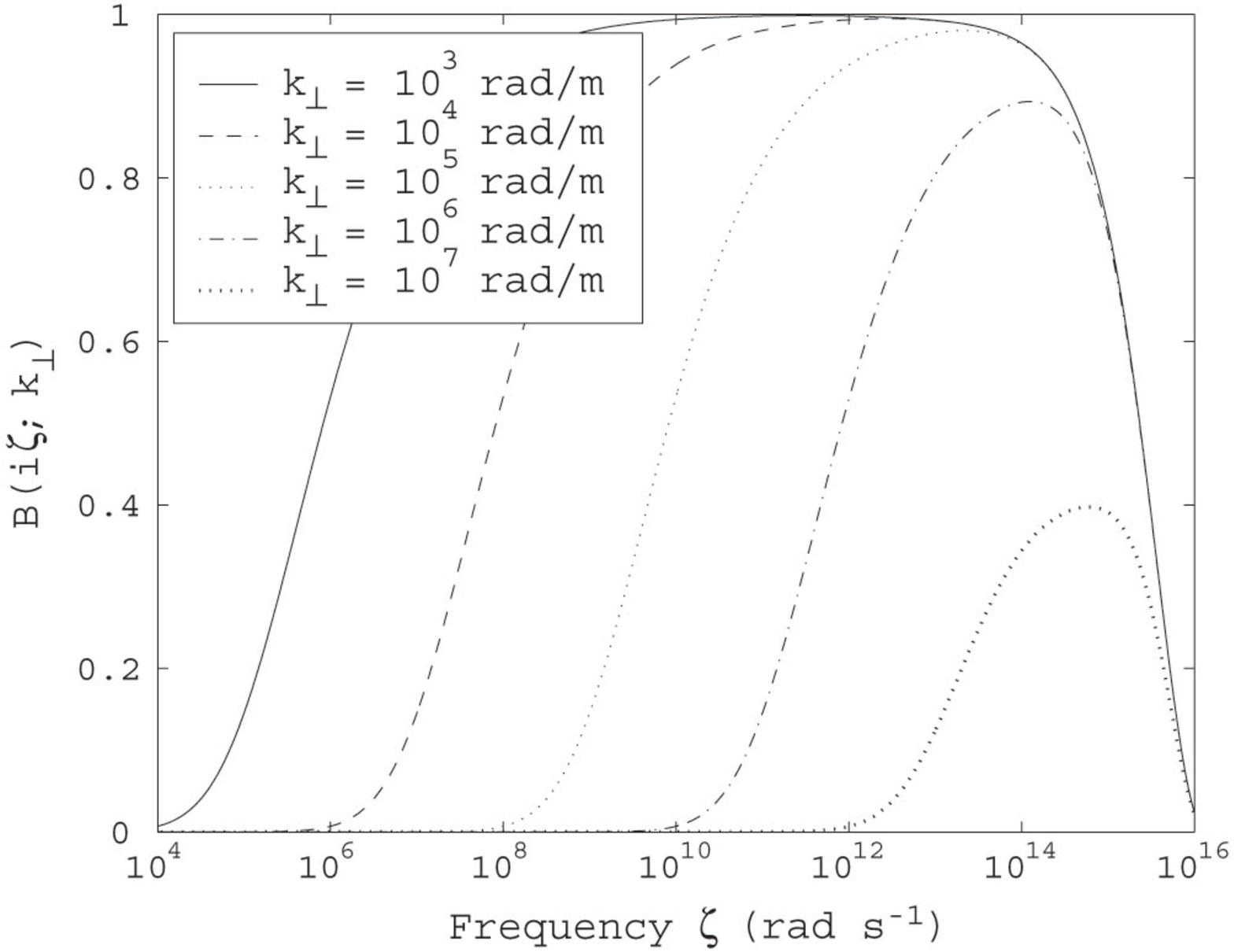}
    \caption{\label{fig2}
Squared reflection coefficients $A$  and $B$  of the metal interfaces
for the TM and TE modes,  as a function of $\zeta/c$ and the transverse
momentum $k_\perp$. a) $A$ for the TM mode, b) $B$ for the TE mode,
c) $B$ for $k_\perp$   and $\zeta$ close to zero.}
  \end{center}
\end{figure}

The vanishing of the zero-frequency mode is connected with the
behaviour of the coefficient $B$ at vanishing frequency. To
illuminate this point, we show in \fref{fig2} both coefficients $A$
and $B$ as a function of imaginary frequency and transverse
momentum $k_\perp$ for an interface between gold and vacuum. In
part (c) of the figure, we see how $B\rightarrow 0$ when $\zeta
\rightarrow 0$ for $k_\perp \neq 0$, whereas $A$ in \fref{fig2}(a)
for the TM mode equals 1 for all $k_\perp$ when $\zeta \rightarrow
0$.

\begin{figure}
\begin{center}
\includegraphics[width=4.5in]{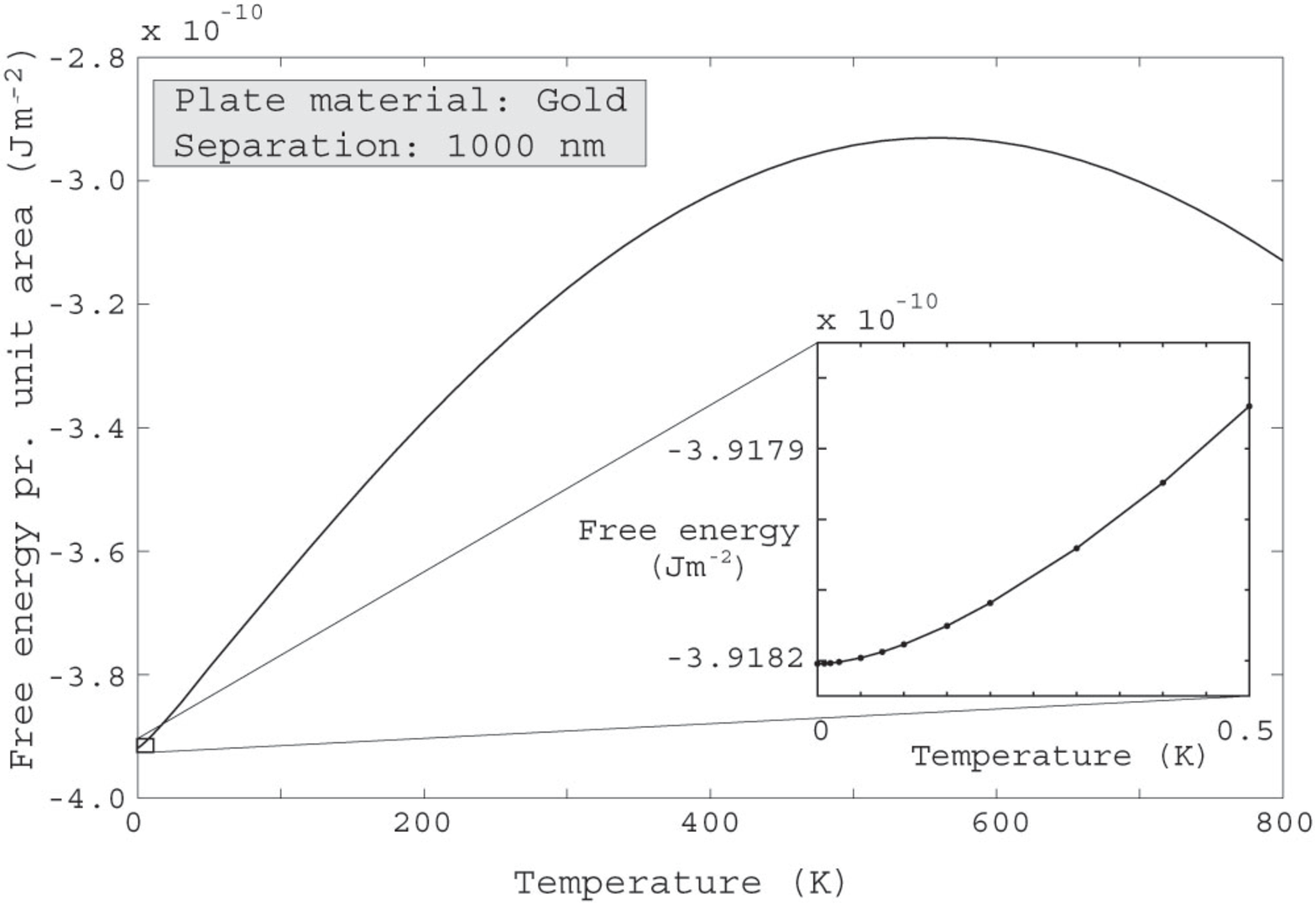}
\caption{\label{fig3}
Numerical evaluation of the free energy (\ref{1}) between
two gold half-spaces as a function of temperature. The inset gives
details for low $T$.}
\end{center}
\end{figure}

By direct numerical integration and lengthy summations independent
of the analytical derivations made in the previous section, we
obtain the free energy numerically. \Fref{fig3} shows the free energy
versus temperature up to 800 K, while the inset shows details of
the parabolic shape close to $T=0$. The figure shows the decrease
of the magnitude of the free energy and thus also the related
decrease of the Casimir force up to a certain temperature. The
inset shows how the slope is horizontal at $T=0$, as predicted.
Thus the entropy at $T=0$ is zero, in accordance with Nernst's theorem.

\section{A more accurate test}

\begin{figure}[h]
\begin{center}
\includegraphics[width=3.5in]{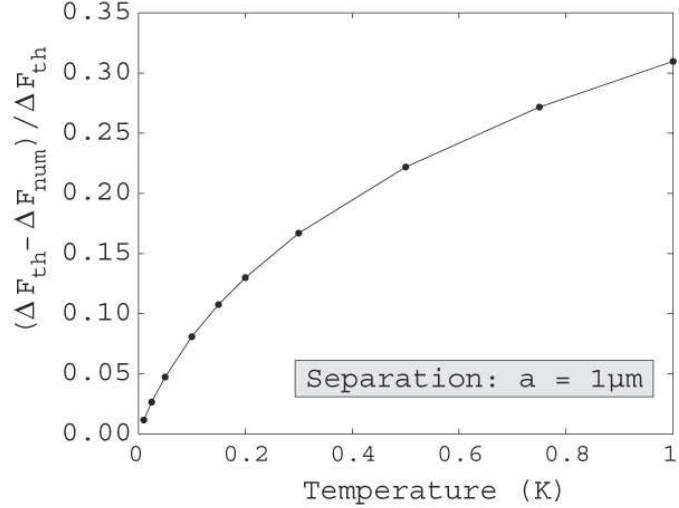}
\end{center}
\caption{\label{fig4}Plot of the ratio $R$ defined in (\ref{25}).}
\end{figure}

Now, there are always uncertainties connected with numerical
calculations. It is possible to make a much more accurate and
sensitive test of the behaviour near $T=0$ in the following way.
Define the quantity $R$ as the relative difference between the
temperature-dependent theoretical free energy $ \Delta
F_{\rm th}^{\rm TE}$, and the temperature-dependent numerical free energy
$ \Delta F_{\rm num}^{\rm TE}$:
\begin{equation}
 R=\frac{\Delta F_{\rm th}^{\rm TE}-\Delta F_{\rm num}^{\rm TE}}{\Delta
F_{\rm th}^{\rm TE}} \label{25}
\end{equation}
Assume for $\Delta F_{\rm th}^{\rm TE}$ the Pad{\'e} approximant form
(\ref{24}), and assume for $\Delta F_{\rm num}^{\rm TE}$ the expansion
\begin{equation}
\Delta F_{\rm num}^{\rm TE}=D_1(T^2-D_2T^{5/2}+D_3T^3+...) \label{26}
\end{equation}
with calculated values for the coefficients $D_1, D_2$ and $D_3$.
Then,
\begin{equation}
R=\frac{C_1-D_1}{C_1}+\frac{D_1}{C_1}(D_2-C_2)T^{1/2}
+\frac{D_1}{C_1}(C_2D_2-D_3)T+...
\label{27}
\end{equation}
If $ C_1=D_1$ and $ C_2=D_2$:
\begin{equation}
 R(T=0)=0, \quad R \propto T, \quad T \rightarrow 0. \label{28}
 \end{equation}
Calculated values of $R$ are plotted in \fref{fig4}. We see that $R$,
when extrapolated, approaches zero linearly with a finite slope.
This demonstrates the  accuracy of the $T^2$ and $T^{5/2}$ terms
in the free energy.

\newpage
\section{Alternative derivation by expansion of $g(m)$}

It may be of interest to mention that, as a variant of the
analytic approach, the dependence of the free energy on $T$ near
$T=0$ can be found by means of complex integration. Start from the
TE expression
\begin{equation}
 \beta F=C{\sum_{m=0}^\infty }{}' \int_{\sqrt{\zeta/D}}^\infty
x\ln (1-B\rme^{-\alpha x})\,\rmd x , \label{29}
\end{equation}
where
\begin{equation}
 C=\frac{\omega_p^2}{\beta
\hbar \nu c^2},\quad \alpha =2a\sqrt{2\pi Cm}, \label{30}
\end{equation}
and expand the logarithm,
\begin{equation}
 \beta F^{TE}=-C\sum_{m=1}^\infty m\int_0^\infty
x\sum_{n=1}^\infty B^n\rme^{-n\alpha x}\,\rmd x. \label{31}
\end{equation}
Now use the formula
\begin{equation}
 \rme^{-n\alpha x}=\frac{1}{2\pi
\rmi}\int_{c-i\infty}^{c+i\infty}\rmd s (n\alpha x)^{-s}\Gamma (s),\quad
4>c>0 \label{32}
 \end{equation}
 and sum over $m$,
 \begin{equation}
 \sum_{m=1}^\infty m^{1-s/2}=\zeta\left(\frac{s}{2}-1\right).
 \label{33}
\end{equation}
 Here $\zeta$ denotes the Riemann zeta function.
Distorting the contour so as to encircle the poles of the $\Gamma$ function
at $s=0, -1, \dots$ then yields the same result \eref{22} as above.

\section{Summary and further remarks}

The main point in our analysis has been to show both analytically
and numerically that the Drude dispersion relation (\ref{6}) does
not run into conflict with basic thermodynamics,  as long as $\nu
\neq 0$ at $T=0$. As we have seen, quite an accurate analysis is
needed for this purpose.  If we instead  had argued in a more
crude way, simply setting the TE coefficient $B_m=0$ for $m=0$ and
keeping all the other coefficients $A_m$ and $B_m$ equal to 1 as
in the modified ideal metal model (MIM), then we would have broken
Nernst's theorem. This issue has been discussed at length in
Refs.~\cite{hoye03} and \cite{hoye06}.

Whether the Drude predictions for the Casimir force are correct or
not is to be decided upon from experiments. A difficulty here is
the inherent uncertainty of theoretical predictions due to the
relatively large spread of published data for the dielectric
permittivity for typical metals such as Au - cf., for instance,
the recent discussions on this point by Pirozhenko {\it et al.}
\cite{pirozhenko06} and Munday and Capasso \cite{munday07}. The
experiment with the highest precision \cite{bezerra06,decca05}
apparently is in disagreement with the Drude model, or any model
satisfying (\ref{11}). It has also been suggested that there are
large thermal effects due to surface roughness \cite{bimonte07a}.
We might  note that the 1\% precision in the dynamic measurement
made by the Purdue group \cite{bezerra06,decca05} is not matched
by the 3\% accuracy of the very recent dynamic experiment reported
in Ref.~\cite{jourdan07}. Our main concern in the present paper,
however, has been to discuss the consistency of this theory.  We
wish to point out that  it would be quite strange if the Drude
relation, proved to be representing permittivity measurements with
great accuracy, should turn out to be inapplicable to explain
Casimir force measurements.  Let us also mention here that an
interesting discussion about the thermal Casimir effect and the
Johnson noise has recently been given by Bimonte \cite{bimonte07},
as a possible theoretical explanation for the discrepancy with
experiment.

The basic assumption for our analysis ought to be re-emphasized.
We assumed the relaxation frequency $\nu$ to be a {\it finite
quantity}, for any value of  $T$. One might here raise the
question: what happens if the metal is a perfect crystal, with no
impurities at all? In such a case $\nu(T=0)=0$, and the formalism
above becomes inapplicable.  (In this we have an opinion different
from the definitive claim of a violation  of the Nernst theorem
given in Refs.~\cite{bezerra04,bezerra06}, for example. See also
Ref.~\cite{intravaia07}.) On basis of the calculation above, we
can thus make no firm statement about the validity of the Nernst
theorem in this special case.

We ought to mention, though, that on physical grounds there are
conceptual difficulties in simply setting $\nu=0$ in the
dispersion relation:

(i)  It would yield a contribution to the Casimir force from the
zero frequency TE mode. This mode is however not a solution of
Maxwell's equations and should therefore not occur. (A more
detailed discussion can be found in Ref.~\cite{hoye03}, and in
Sect.~III in Ref.~\cite{hoye01}.)

(ii) Introducing a zero TE mode for perfect crystals would imply
that such a medium would behave differently from a real metal when
taking the limit $\nu=0$. This would create a discontinuity in
behaviour that we find unphysical.

\vspace{0.3cm}

There are additional physical effects that we have not taken into
account above:

(1)  One such effect is {\it spatial dispersion}
\cite{sernelius05}, implying that  the wave vector $\bf k$ is
present in the dispersion relation. Then $\varepsilon
=\varepsilon(\omega, k)$ would become finite for finite $ k$. Only
the special case $\varepsilon(0, 0)$ would be infinite, and it
would not appear natural that this "measure zero" case should
yield a finite contribution to the Casimir force.\footnote{It
could be mentioned here that the contrary view has been expressed
in Ref.~\cite{klimchitskaya07c}.}

(2)  Another effect that could have been taken into account is the
{\it anomalous  skin effect} \cite{esquivel04,svetovoy05}. This
effect occurs when the mean free path in the metal becomes much
larger than the field penetration depth near $ T=0$. Again, no
contribution to the Casimir force is found from the zero TE mode,
and the Nernst theorem is satisfied.

Finally, we refer to the very recent microscopic theory of the
Casimir force at large separations, i.e.~the classical limit,
 using statistical mechanics
\cite{buenzli07} - cf. also \cite{buenzli07a} and further
references therein. These authors make use of a joint functional
representation of both matter and field, enabling them to
integrate out the field degrees of freedom entirely. Important in
our context is that they find the TE modes not to contribute in
this regime, and that  the Casimir surface pressure is
\begin{equation}
P=-\frac{\zeta(3)kT}{8\pi a^3}, \quad a\rightarrow \infty.
\label{33a}
\end{equation}
This is precisely as predicted by the Drude model in the same
limit. This conclusion is further supported by Svetovoy's recent
demonstration \cite{svetovoy07} of the cancellation between TE
evanescent wave (EW) and propagating wave (PW) contributions for
large distances, yielding Eq.~(\ref{33a}),  while at short
distances the TE EW dominates for the force between two metal
plates or between a metal plate and a dielectric plate, resulting
in a linear temperature term in the force.

\ack

We thank Astrid Lambrecht for sending us the very useful
permittivity data.  The work of KAM is supported by grants
from the US Department of Energy and the US National Science
Foundation.


\section*{References}
\begin{thebibliography}{99}
\bibitem{hoye07}
H{\o}ye J S, Brevik I, Ellingsen S A and Aarseth J B 2007 {\it
Phys. Rev. E} {\bf 75} 051127
\bibitem{brevik06}
Brevik I, Ellingsen S A, and Milton K A 2006 {\it New J. Phys.}
{\bf 8} 236
\bibitem{brevik06a}
Brevik I and Aarseth J B 2006 {\it J. Phys. A: Math. Gen.} {\bf
39} 6589
\bibitem{brevik05}
Brevik I, Aarseth J B, H{\o}ye J S and Milton K A 2005 {\it Phys.
Rev. E} {\bf 71} 056101
\bibitem{brevik04}
Brevik I, Aarseth J B, H{\o}ye J S and Milton K A 2004 {\it Proc.
6th Workshop on Quantum Field Theory under the Influence of
External Conditions}  ed K A Milton (Paramus, NJ: Rinton Press) p
54 ({\it Preprint} quant-ph/0311094)
\bibitem{hoye03}
H{\o}ye J S, Brevik I, Aarseth J B and Milton K A 2003 {\it Phys.
Rev. E} {\bf 67} 056116
\bibitem{klimchitskaya07}
Klimchitskaya G L and Mostepanenko V M 2007 {\it Preprint}
quant-ph/0703214
\bibitem{khoshenevisan79}
Khoshenevisan M, Pratt Jr W P, Schroeder P A and Steenwyk S D 1979
{\it Phys. Rev. B} {\bf 19} 3873
\bibitem{bostrom00}
Bostr{\"o}m M and Sernelius Bo E 2000 {\it Phys. Rev. Lett.} {\bf
84} 4757
\bibitem{jancovici00}
Jancovici B and {\v S}amaj L 2005{\it Europhys. Lett.} {\bf 72}
35
\bibitem{buenzli05}
Buenzli P R and Martin Ph A 2005 {\it Europhys. Lett.} {\bf 72} 42
\bibitem{geyer07}
Geyer B, Klimchitskaya G L and Mostepanenko V M 2007 {\it
Preprint} arXiv:0710.0254 [quant-ph]
\bibitem{chen07}
Chen F, Klimchitskaya G L, Mostepanenko V M and Mohideen U 2007
{\it Optics Express} {\bf 15} No.~8
\bibitem{klimchitskaya07a}
Klimchitskaya G L, Mohideen U and Mostepanenko V M 2007 {\it J.
Phys. A: Math Theor.} {\bf 40} F339
\bibitem{bezerra06}
Bezerra V B, Decca R S, Fischbach E, Geyer B, Klimchitskaya G L,
Krause D E, L{\'o}pez D, Mostepanenko V M and Romero C 2006 {\it
Phys. Rev. E} {\bf 73} 028101
\bibitem{decca05}
Decca R S, L{\'o}pez D, Fischbach E, Klimchitskaya G L, Krause D E
and Mostepanenko V M 2005 {\it Ann. Phys. (NY)} {\bf 318} 37
\bibitem{bezerra04}
Bezerra V B, Klimchitskaya G L, Mostepanenko V M and Romero C 2004
{\it Phys. Rev. A} {\bf 69} 022119
\bibitem{hoye06}
H{\o}ye J S, Brevik I, Aarseth J B and Milton K A 2006 {\it J.
Phys. A: Math. Gen.} {\bf 39} 6031
\bibitem{pirozhenko06}
Pirozhenko I, Lambrecht A and Svetovoy, V B 2006 {\it New J.
Phys.} {\bf 8} 238
\bibitem{munday07}
Munday J N and Capasso F 2007 {\it Preprint} arXiv:0711.2437
[quant-ph]
\bibitem{bimonte07a}
Bimonte G 2007 {\it Preprint} arXiv:0711.0278v2 [quant-ph]
\bibitem{jourdan07}
Jourdan G, Lambrecht A, Comin, F and Chevrier, J 2007 {\it
Preprint} arXiv:0712.1767 [physics.gen-ph]
\bibitem{bimonte07}
Bimonte G 2007 {\it New J. Phys.}  {\bf 9} 281
\bibitem{intravaia07}
Intravaia F and Henkel C {\it Preprint} arXiv:0710.4915
[quant-ph]; {\it J. Phys. A: Math. Theor.}, to appear
\bibitem{hoye01}
H{\o}ye J S, Brevik I and Aarseth J B 2001 {\it Phys. Rev. E} {\bf
63} 051101
\bibitem{sernelius05}
Sernelius Bo E 2005 {\it Phys. Rev. B} {\bf 71} 235114
\bibitem{klimchitskaya07c}
Klimchitskaya G L and Mostepanenko V M 2007 {\it Phys. Rev. B}
{\bf 75} 036101
\bibitem{esquivel04}
Esquivel R and Svetovoy V B 2004 {\it Phys. Rev. A} {\bf 69}
062102
\bibitem{svetovoy05}
Svetovoy V B and Esquivel R 2005 {\it Phys. Rev. E} {\bf 72}
036113
\bibitem{buenzli07}
Buenzli P R and Martin Ph A 2007 {\it Preprint} arXiv:0709.4194
[quant-ph]
\bibitem{buenzli07a}
Buenzli PR, Martin Ph A and Ryser M D 2007 {\it Phys. Rev. E} {\bf
75} 041125
\bibitem{svetovoy07}
Svetovoy V B {\it Preprint} arXiv:0711.0841 [quant-ph]
\end{thebibliography}
\end{document}